**The impact of imbalanced training data on machine learning for author name disambiguation**

**Author:** Jinseok Kim and Jenna Kim

**Author Information:**

Jinseok Kim

Institute for Research on Innovation and Science, Survey Research Center, Institute for Social Research, University of Michigan
330 Packard Street, Ann Arbor, MI U.S.A. 48104
734-763-4994
jinseokk@umich.edu; jinseok.academic@gmail.com
ORCID ID: 0000-0001-6481-2065

Jenna Kim

School of Information Studies, Syracuse University
343 Hinds Hall, Syracuse, NY U.S.A. 13210
jkim141@syr.edu

Abstract

In supervised machine learning for author name disambiguation, negative training data are often dominantly larger than positive training data. This paper examines how the ratios of negative to positive training data can affect the performance of machine learning algorithms to disambiguate author names in bibliographic records. On multiple labeled datasets, three classifiers – Logistic Regression, Naïve Bayes, and Random Forest – are trained through representative features such as coauthor names, and title words extracted from the same training data but with various positive-negative training data ratios. Results show that increasing negative training data can improve disambiguation performance but with a few percent of performance gains and sometimes degrade it. Logistic Regression and Naïve Bayes learn optimal disambiguation models even with a base ratio (1:1) of positive and negative training data. Also, the performance improvement by Random Forest tends to quickly saturate roughly after 1:10 ~ 1:15. These findings imply that contrary to the common practice using all training data, name disambiguation algorithms can be trained using part of negative training data without degrading much disambiguation performance while increasing computational efficiency. This study calls for more attention from author name disambiguation scholars to methods for machine learning from imbalanced data.

Keywords: author name disambiguation; negative training data; imbalanced training data; supervised machine learning

Introduction

Author name ambiguity has been a daunting challenge to scholars who mine bibliographic data for scientific knowledge (Garfield, 1969). Many scholars have solved the problem using simple heuristics such as forename-initial-based matching (i.e., two author names are regarded to refer to the same author if they match on a forename initial(s) and full surname) (e.g., Barabási et al., 2002; Newman, 2001). As noted in several recent studies, these heuristics can merge and split author entities (e.g., two authors with the same forename initials and full surname can be regarded as a single entity), leading to inaccurate understanding of bibliographic data (e.g., Fegley & Torvik, 2013; J. Kim & Diesner, 2016).

A proactive approach to the name ambiguity problem is to use computing algorithms to distinguish author entities. A variety of algorithm-based disambiguation methods has been developed by computer and information scientists (Smalheiser & Torvik, 2009). Among them, supervised machine learning has been reported to produce decent to highly accurate disambiguation results, although its performance can vary depending on characteristics of target bibliographic data (e.g., small, medium, and large data with different levels of name ambiguity) and types of algorithms (Ferreira, Goncalves, & Laender, 2012).

Regardless of algorithmic variations, supervised machine learning for author name disambiguation typically requires labeled training data in which author identification tags (i.e., labels) are assigned to author name instances by, in most cases, laborious manual identity checking (Muller, Reitz, & Roy, 2017). Pairs of name instances with the same labels constitute a "positive" training dataset, while pairs with different labels construct a "negative" training dataset. Then, name instances within positive and negative training datasets are compared pairwisely for calculating their similarity across various features such as coauthor names, affiliation, paper title, and publication venue. The resulting similarity profiles (often vectors of similarity scores) between comparison pairs are fed into machine learning algorithms so that the algorithms can learn disambiguation patterns to decide whether any pair of name instances under test refers to the same author or not.

This study is motivated by the observation that in many labeled data for author name disambiguation, positive and negative training data are often imbalanced. This situation is illustrated in Table 1. Let's assume that five name instances (#1 ~ #5) require disambiguation in Table 1, where each instance is labeled with one of four distinct authors (A, B, C, and D). Among ten possible pairwise comparison pairs, only one positive pair (Instance 1 and Instance 2 with the same label A) exists, leaving nine pairs as negative sets. Such imbalance can increase dramatically if the number of names to disambiguate is large while those names are associated with many distinct authors.

*Table 1: An Illustration of Positive and Negative Training Data Imbalance in Author Name Disambiguation*

| Name Instance # | Name String | Author Label | Pairs | |
|---|---|---|---|---|
| | | | Positive | Negative |
| 1 | J. Kim | A | 1-2 | 1-3, 1-4, 1-5, 2-3, 2-4, 2-5, 3-4, 3-5, 4-5 |
| 2 | J. Kim | A | | |
| 3 | J. Kim | B | | |
| 4 | J. Kim | C | | |
| 5 | J. Kim | D | | |

This positive and negative training data imbalance can be observed in many labeled data generated by collating most ambiguous name instances (Muller et al., 2017). In a study of blocking methods for author name disambiguation (Kim, Sefid, & Giles, 2017), for example, its labeled data contained 3,964 name instances of 214 distinct authors who are associated with 10 ambiguous names (e.g., S Kim, C lee, J Smith, etc.). Among a total of 7.8M comparison pairs, only 51,052 (0.65%) pairs were positive training pairs.

As such, negative training data can be more abundant than positive training data in supervised machine learning for author name disambiguation, consuming much computation time and resources. But how such prevalence of negative training data can affect the performance of author name disambiguation algorithms has been insufficiently discussed. To contribute to the discussion, this study examines the impact of positive and negative training data imbalance on machine learning for disambiguating author names in publication records. For this purpose, this study compares the performances of three machine learning algorithms – Logistic Regression, Naïve Bayes, and Random Forest – that are tested on different positive-negative training data ratios. By doing so, this study aims to help scholars determine the optimal positive-negative training data ratios to yield good disambiguation results with increased computational efficiency. In following section, prior work on imbalanced training data is presented to contextualize this study.

## Related Work

In machine learning research, the problem of imbalanced data has continued to receive scholarly attention (Bickel, Bruckner, & Scheffer, 2009; He & Garcia, 2009; Shimodaira, 2000). But most studies have been focused on addressing the imbalance across training and test data, resulting in a variety of sampling methods to improve the performance of machine learning models trained on imbalanced data. Meanwhile, a few studies have investigated the imbalanced training data problem for text classification tasks. For example, arguing that negative training data do not improve much machine learning performance and sometimes degrade it, several scholars have proposed the PU learning model that discards negative training data and relies only positive (P) and unlabeled (U) training data (Li, Liu, & Ng, 2010; Liu, Dai, Li, Lee, & Yu, 2003).

In bio- and chemical informatics, the *training* data imbalance has been actively studied because negative training data tend to be dominant while positive training data can be scarce (e.g., non-cancer vs. cancer patients) (Chawla, Bowyer, Hall, & Kegelmeyer, 2002; Woods et al., 1993). Some studies found that increasing the ratio of negative to positive training data improves machine learning performance but such improvement was negligible after the positive-to-negative training data ratios of around 1:10 (Heikamp & Bajorath, 2013; Kurczab, Smusz, & Bojarski, 2014). In addition, Kurczab et al. (2014) reported that with increased training data, recall tends to suffer from degraded performance while precision is improved.

Scholars in author name disambiguation have also faced the training data imbalance problem. Typically, the number of comparison pairs in a block (i.e., a group of name instances to be compared with one another for author name disambiguation) increases quadratically with the block size. Among them, the number of positive pairs of name instances (i.e., referring to the same authors) can be very small, while the number of negative pairs (i.e., referring to different authors) can be large. This situation was illustrated in Table 1 using a simple scenario. Facing this problem, many scholars have used various methods to partition name instances into blocks so that only name instances that are likely to refer to the same authors are collected in the same blocks, thereby reducing the number of non-matching (negative) comparison pairs (for a recent review, see Kim et al. (2017)). Once blocks are generated, however, the common practice in author name disambiguation research is to utilize all comparison pairs in each block or in a rare case, uniformly sample training pairs regardless of whether they are positive or negative ones (e.g., Han, Giles, Zha, Li, & Tsioutsiouliklis, 2004).

So, the impact of imbalanced positive and negative training data on machine learning is still an under-researched topic for author name disambiguation. Studying this topic can provide methodological insights for future disambiguation research and its application to disambiguating author names in growing digital libraries. As reported in aforesaid chemical informatics studies, for example, part of negative training data may be used to train name disambiguation algorithms with negligible performance degradation on test data, improving further computational efficiency in conjunction with well-designed blocking schemes. In contrast, however, partial use of negative data may not be so useful. According to Levin, Krawczyk, Bethard, and Jurafsky (2012), for example, reducing negative training data size leads to a poor performance while using all negative data produces the best outcome.

Therefore, this study takes the impact analysis approach of aforesaid studies such as Kurczab et al. (2014) and Levin et al. (2012) to obtain a better understanding of how the imbalance of positive and negative training data can affect algorithmic author name disambiguation. Specifically, this study empirically tests the performances of three representative machine learning algorithms for author name disambiguation – Logistic Regression, Naïve Bayes, and Random Forest – that are trained on various labeled data in which positive-negative data ratios are incrementally increased from the equal ratio. Details of labeled data for analysis and machine learning settings are provided in the following section.

## Methodology

Data

*GILES*: For the impact analysis of imbalanced training data, this study uses three representative labeled data for author name disambiguation (Muller et al., 2017). The first data[1] were generated by Dr. Giles's research lab at the Pennsylvania State University (Han et al., 2004; Han, Zha, & Giles, 2005). The GILES (hereafter) data have been widely used for training various author name disambiguation algorithms (e.g., Cota, Ferreira, Nascimento, Goncalves, & Laender, 2010; Santana, Goncalves, Laender, & Ferreira, 2015). The data consist of 8,453 highly ambiguous name instances (e.g., A. Gupta, S. Lee, and J. Smith) and their associated publication records that are gathered from the computing research library DBLP and webpages of authors. Distinct author labels were assigned to name instances manually by human coders. Recently, several studies have noted that the original GILES data contain duplicate and erroneous records (Muller et al., 2017; Santana et al., 2015; Shin, Kim, Choi, & Kim, 2014). So, following Kim (2018), this study removed duplicate records in the original GILES data. For error correction (e.g., missing coauthor names), records in the de-duplicated GILES data were updated by publication records in DBLP[2] that were matched to GILES records through the comparison of author name, year, title, and venue. If a record in GILES has no match in DBLP, it was excluded from analysis. This cleaning process resulted in a total of 5,018 name instances and their associated records (59% of the original GILES data) labeled for 480 distinct authors[3].

*KANG*: The second labeled data (KANG hereafter)[4] were created by Korean scholars (Kang, Kim, Lee, Jung, & You, 2011) and have been used in several disambiguation studies (e.g., Santana et al., 2015). The KANG data contain 41,673 author name instances and their publication records extracted from DBLP. Labels of 6,921 unique authors were assigned to each name instance through a semi-manual disambiguation by triangulating Google search results and human inspection.

---

[1] http://clgiles.ist.psu.edu/data/nameset_author-disamb.tar.zip
[2] dblp.org/xml/release/dblp-2017-09-03.xml.gz
[3] Available at https://figshare.com/articles/DBLP-derived_labeled_data_for_author_name_disambiguation/6840281
[4] http://www.lbd.dcc.ufmg.br/lbd/collections/disambiguation/DBLP.tar.gz/at_download/file

*TANG*: Another labeled data (TANG hereafter)[5] were constructed by Chinese scholars led by Dr. Tang at the Tsinghua University in China to train disambiguation algorithms for the computing research digital library AMiner (Tang, Fong, Wang, & Zhang, 2012; Wang, Tang, Cheng, & Yu, 2011). Dr. Tang's team gathered 7,528 name instances associated with 110 ambiguous full names and manually disambiguated them, assigning 1,546 unique author labels.

Machine Learning Settings

*Overview*: Broadly speaking, there are two approaches to author name disambiguation: author clustering and author assignment (Ferreira, Goncalves, & Laender, 2012). This study uses the author clustering method which typically consists of two phases - (1) classification of match/non-match between pairs of name instances and (2) clustering name instances based on the classification decision. An author clustering method first decides which pairs of these name instances are likely or unlikely to refer to the same author by comparing information extracted from features such coauthor names. During this process, a classification algorithm is used to learn the match/non-match patterns from training data and predict match/non-match of newly seen pairs in test data. As a result of this classification, we have pairs of name instances that refer to same authors and pairs to refer to different authors. Next step is to collate name instances that refer to same authors using these pairwise decisions, which is called "clustering." Here, a problem arises when dyadic match/non-match decisions can contradict each other. Let's take an example of Instance A = Instance B, Instance B = Instance C, and Instance A ≠ Instance C. According to a transitivity rule, Instance A = Instance C is logical but algorithms can often produce this kind of contradictory decisions because they conduct prediction only at a pair level. To resolve this problem, many disambiguation studies use supervised or unsupervised clustering algorithms to detect optimal groups of name instances that are likely to refer to same authors after the pairwise classification decisions. For this, specifically, the pairwise classification decisions by algorithms are output as similarity scores usually between 0 and 1 calculated across features, instead match/non-match binary decisions. Then, clustering algorithms group name instances based on these similarity scores. Number of resulting clusters (= number of distinct authors) can vary depending on the threshold of similarity scores. If a truth number of clusters is given, clustering algorithms will find the best threshold to produce that number of clusters. In this paper, the truth number of clusters is given by labeled test data.

*Training Data*: In many author name disambiguation studies, name instances that share the first forename initial and full surname are collated into a block (i.e., blocking) as a pre-disambiguation step to reduce the amount of pairwise comparison pairs (e.g., Han et al., 2004; Levin et al., 2012; Santana et al., 2015; Wang et al., 2011). Following this common practice, this study conducted algorithmic disambiguation on names in the same block. Name instances and their associated publication records in each block were randomly divided into two subsets – training data (50%) and test data (50%). Then, positive (i.e., with identical labels) and negative (i.e., with different labels) pairs of name instances were generated from the per-block training data with different positive-negative pair ratios. For this, specifically, the number of positive pairs was first counted. Then, among all possible negative pairs, a subset of them was randomly selected to make the ratios of negative to positive training pairs increased incrementally from 1:1 up to 1:$R$, where $R$ is the maximum ratio that equals to the (round-down) integer of the total of negative training pairs divided by the total of all positive training pairs.

*Feature Selection*: In author name disambiguation research, many features have been engineered and tested to find ones that contribute most to disambiguation performance (Tang & Walsh, 2010; Wang et al., 2012). This study aims to show how the different ratios of negative to positive training data may affect performances of disambiguation algorithms. A challenge is that if we use many features, we cannot distinguish the impact of different positive-negative training data ratios from the impact of feature effectiveness. So, we tried to select a minimum set of features – coauthor names and title words – which are commonly used in most disambiguation studies and have been found to be effective in disambiguating names (Ferreira et al., 2012; Schulz, 2016; Wang et al., 2012). Another reason is that these two features are available across all labeled datasets used in this study, while other features such as affiliation, journal names, and references are recorded in some data but not in another. To run disambiguation tests fairly on all labeled datasets, therefore, two commonly used features –coauthor names and title words – that are associated with name instances in training data were chosen to generate a similarity score vector between a pair of name instances. Across features, all text strings were lower-cased and special characters were encoded into ASCII. Non-alphanumeric characters were replaced by spaces except commas because they separate the forename of an author name from its surname. Each title word was stemmed by the Porter's Stemmer (Porter, 1980)[6] after common English words such as pronouns and prepositions were stop-listed[7]. All (co)author names were converted into the format of first forename initial and full surname (e.g., J. Wang) as KANG and TANG record many author names in full name while GILES records the majority of names in the format of full surname and initialized-forename. This pre-processing of author names was conducted to reduce the confounding impact of name string on disambiguation performance other than positive-negative training data ratios (Han et al., 2005; Louppe, Al-Natsheh, Susik, & Maguire, 2016). Similarity scores between a pair of name instances were calculated by the cosine similarity of TF-IDF for 2, 3, and 4-grams over each feature, following the practice of several studies (e.g., Han et al., 2005; Levin et al., 2012; Louppe et al., 2016; Santana et al., 2015; Treeratpituk & Giles, 2009).

---

[5] http://arnetminer.org/lab-datasets/disambiguation/rich-author-disambiguation-data.zip

*Classifiers and Clustering*: The resulting pairwise similarity scores for positive and negative training pairs were used for training three machine learning algorithms – Logistic Regression, Naïve Bayes, and Random Forest[8] – that represent base classifiers frequently run in author disambiguation research (e.g., Hui Han, Xu, Zha, & Giles, 2005; Levin et al., 2012; Santana et al., 2015; Torvik & Smalheiser, 2009; Treeratpituk & Giles, 2009; Wang et al., 2012). The trained models by these algorithms were applied to disambiguating author name instances in test data. Specifically, name instances in test data were pairwisely compared for a similarity profile in the same way name instances in training data were compared. Then, each pair of name instances was assigned a probability score to refer to the same author based on the disambiguation model learned from training data by each algorithm. Using the probability score between a pair as a proxy of similarity distance between them (*higher* score means *closer* distance between a pair), the hierarchical agglomerative clustering algorithm grouped name instances that belong to the same author into a cluster. A threshold distance to decide the number of distinct clusters in test data was determined by trying various threshold values to maximize the clustering accuracy which was evaluated on the labels associated with name instanced in test data (Louppe et al., 2016)[9].

Accuracy Measure

A suite of *B*-Cubed ($B^3$) metrics (Bagga & Baldwin, 1998) was used to calculate disambiguation accuracy. Three parts of this measure – $B^3$ Precision (*b*P), $B^3$ Recall (*b*R), and $B^3$ F1 (*b*F1) – are defined as follows:

---

[6] Codes by Martin Porter are available at https://tartarus.org/martin/PorterStemmer/
[7] https://github.com/stanfordnlp/CoreNLP/blob/master/data/edu/stanford/nlp/patterns/surface/stopwords.txt
[8] Three classifiers were implemented by Scikit-Learn Python packages at http://scikit-learn.org/stable/index.html
[9] Substantial part of the training and test procedure was conducted by modifying Python codes generously shared by Louppe et al. (2016). The original codes are available at https://github.com/glouppe/paper-author-disambiguation

$$bP = \frac{1}{N} \times \sum_{i \in L} \frac{|C_D(i) \cap C_L(i)|}{|C_D(i)|} \quad (1)$$

$$bR = \frac{1}{N} \times \sum_{i \in L} \frac{|C_D(i) \cap C_L(i)|}{|C_L(i)|} \quad (2)$$

$$bF1 = \frac{2 \times bP \times bR}{bP + bR} \quad (3)$$

Here, $C_D(i)$ means a cluster of name instances that contains the name instance $i$ and is decided to refer to the same author as a result of algorithmic disambiguation, while $C_L(i)$ means a cluster of name instances that contains the name instance $i$ and refer to the same author in labeled data. $N$ is the number of name instances in labeled data (L).

The $B^3$ metrics and its variations have been used in many entity resolution studies as well as author name disambiguation research (Ferreira, Veloso, Goncalves, & Laender, 2014; Levin et al., 2012; Louppe et al., 2016; Menestrina, Whang, & Garcia-Molina, 2010). The $B^3$ metrics were chosen over another frequently used *pairwise*-F metrics because the former calculates disambiguation accuracy at an instance level while the latter excludes an instance with no comparable pair from calculation. This can impact the disambiguation evaluation for data in which many disambiguated instances form singleton clusters. In addition, as the number of comparison pairs increases in a quadratic way with the size of instances in a cluster, the results of the *pairwise*-F calculation can be biased towards large clusters, while by the instance-based B-Cubed measure, clusters affect calculation linearly with their size (Levin et al., 2012; Louppe et al., 2016).

Results

Per-Block Analysis

To observe the impact of imbalanced positive-negative training data on the disambiguation performances of three classification algorithms, name blocks in GILES were used for the training and test simulation per positive-negative training data ratio. A summary of training data in GILES is reported per block in Table 2. The GILES data contain a total of 14 blocks: A. Gupta, A. Kumar, C. Chen, D. Johnson, J. Lee, J. Martin, J. Robinson, J. Smith, K. Tanaka, M. Brown, M. Jones, M. Miller, S. Lee, and Y. Chen. For the purpose of simplicity, four blocks with $R \leq 5$ were excluded from analysis: A. Kumar ($R = 3$), D. Johnson ($R = 2$), M. Miller ($R = 3$), and K. Tanaka ($R = 2$). The table shows that negative training pairs are more abundant than positive ones across blocks (see the "No. of Training Pairs" column in Table 2). The maximum ratios ($R$) of negative to positive training pairs range from 1:7 (M. Brown and M. Jones) to 1:47 (J. Lee).

*Table 2: Summary of Training Pairs per Block in GILES Data (R represents the maximum ratio of negative to positive training pairs)*

| Block | No. of Instances (train + test) | No. of Authors (train + test) | No. of Training Pairs | | | 1…R |
| --- | --- | --- | --- | --- | --- | --- |
| | | | Total | Positive | Negative | |
| A. Gupta | 470 | 27 | 27,495 | 2,936 | 24,559 | 1…8 |
| C. Chen | 475 | 61 | 27,966 | 903 | 27,063 | 1…29 |
| J. Lee | 855 | 100 | 90,525 | 1,853 | 88,672 | 1…47 |
| J. Martin | 94 | 16 | 1,081 | 112 | 969 | 1…8 |
| J. Robinson | 142 | 12 | 2,485 | 347 | 2,138 | 1…6 |
| J. Smith | 479 | 30 | 28,441 | 3,032 | 25,409 | 1…8 |
| M. Brown | 109 | 13 | 1,431 | 170 | 1,261 | 1…7 |
| M. Jones | 166 | 13 | 3,403 | 392 | 3,011 | 1…7 |
| S. Lee | 960 | 86 | 114,960 | 5,027 | 109,933 | 1…21 |
| Y. Chen | 547 | 71 | 37,128 | 929 | 36,231 | 1…38 |

The train-and-test procedure detailed in the "Machine Learning Settings" section was repeated 10 times for each positive-negative training data ratio per block and accuracy scores were averaged for report. The average $B^3$ precision ($b$P), recall ($b$R), and harmonic mean ($b$F1) scores of three classifiers per positive-negative data ratio are presented in Figure 1 (C. Chen, J. Lee, S. Lee, and Y. Chen) and Figure 2 (A. Gupta, J. Martin, J. Robinson, J. Smith, M. Brown, and M. Jones). In subfigures of Figure 1 and Figure 2, positive-negative training data ratios (1 up to $R$) are denoted on $x$-axes, while mean accuracy scores are on $y$-axes. A note is that endpoints of trend lines (i.e., $R + 1$ on $x$-axes) represent accuracy scores when all negative training data are used for machine learning.

An overall trend in both Figure 1 and Figure 2 is that increasing the ratios of negative training data improved the precision ($b$P) scores by three classifiers in many blocks. This precision improvement is visually represented by plots moving slightly toward the upper and right corners in each "Precision" subfigure. Such improvement became, however, less pronounced with larger negative training data, which is depicted by the flattened accuracy plots. In addition, some author name blocks such as J. Lee (Figure 1), A. Gupta (Figure 2), and M. Jones (Figure 2) showed degraded performances by Logistic Regression and Naïve Bayes as the negative training data size increased.

Likewise, the recall ($b$R) plots showed mixed trends depending on name blocks and classifiers. In all four blocks in Figure 1, for example, performance gains by the increased negative training data were clearly observed for Random Forest over the positive-negative ratio range of roughly 1:1 ~ 1:15. But $b$R trends by Logistic Regression and Naïve Bayes tended to move downward or flattened as their positive-negative training data ratios increased.

Compared to pronounced variations in precision and recall, their harmonic mean ($b$F1) did not show much score variations across name blocks and classifiers. The $b$F1 plots for Logistic Regression (LR) and Naïve Bayes (NB) in Figure 1 moved rightward horizontally without much fluctuation. The $b$F1 plots for Random Forest (RF) showed slightly rising trends until the ratios of negative to positive reached roughly 1:10 ~ 1:15 but almost flattened beyond those ratios. For small blocks in Figure 2, a similar not-so-much wavering pattern was observed for $b$F1 plots by LR and NB, while those by RF showed a mixture of up and down movements. This indicates that for each classifier, precision gains from the increased negative training data were often offset by recall losses.

The aforesaid observations indicate that part of negative training data can be effective in training machine learning algorithms for author name disambiguation. For large blocks in Figure 1, specifically, the performance gains ($b$F1) by Random Forest tended to be substantial as the negative data size increased but this improvement reached a saturation point at around $R = 10$ ~ 15. Regarding Naïve Bayes and Logistic Regression classifiers, however, the added performance gains by the increased negative training data were negligible: their $b$F1 plots were flat across most positive-negative data ratios. Even for small blocks in Figure 2, the change of negative training data ratios did not produce much enhanced results by Logistic Regression and Naïve Bayes algorithms in terms of $b$F1, while Random Forest produced slightly improved performance with larger negative training data. This means that two algorithms – Logistic

Regression and Naïve Bayes – produced optimal models very quickly using small part of negative training data, while Random Forest continued to improve models from increased negative training data. Another noteworthy observation is that adding negative training data can be detrimental to disambiguation performances depending on the types of accuracy measure (precision versus recall) and classifiers, as illustrated by J. Lee in Figure 1 (see LR and NB for precision) and most blocks in Figure 2 (see NB and RF for recall). For example, the J. Lee block showed decreases in all $B^3$ scores occasionally by Random Forest as the negative training data size increased.

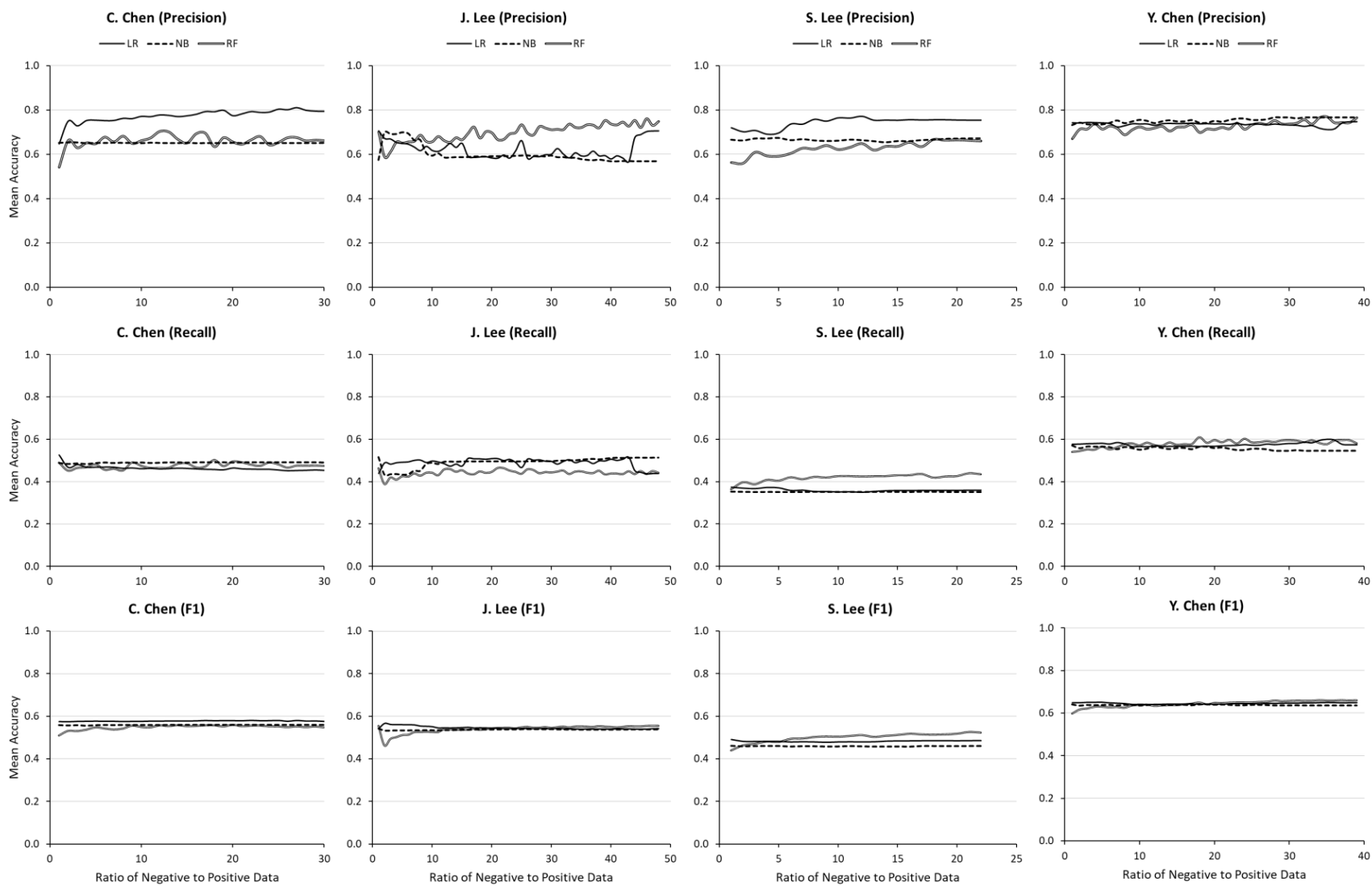

Figure 1: Trends of Mean Accuracy of Author Name Disambiguation per Positive-Negative Training Pair Ratio for Four Blocks in GILES Data (x-axes denote positive-negative training pair ratios from 1:1 to 1:R while y-axes denote mean accuracy scores of B-cubed precision, recall, and F1 measured on test data)

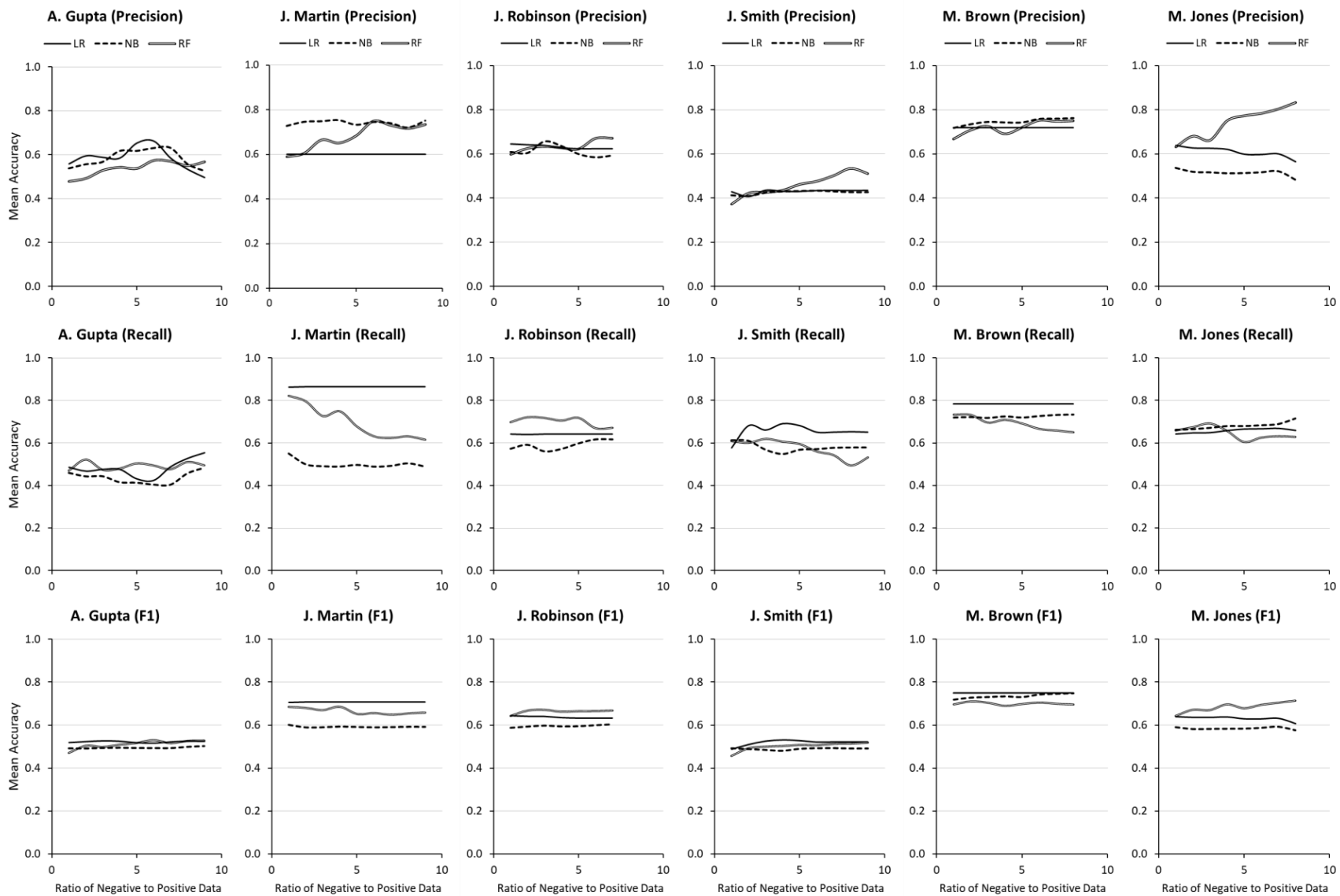

*Figure 2: Trends of Mean Accuracy of Author Name Disambiguation per Positive-Negative Training Pair Ratio for Six Blocks in GILES Data (x-axes denote positive-negative training pair ratios from 1:1 to 1:R while y-axes denote mean accuracy scores of B-cubed precision, recall, and F1 measured on test data)*

Cross-Data Comparison

The idea that part of negative training data may be effective to train name disambiguation algorithms was tested on three labeled datasets – GILES, KANG, and TANG – applying the same train-and-test procedure detailed in the "Machine Learning Settings" section. For this purpose, especially, only blocks containing 100 or more name instances were selected from original KANG and TANG datasets to be consistent with the GILES data in which all blocks have almost 100 or more instances.

Based on the aforementioned observations from 10 blocks in GILES, three bounds of $R$ – 1:1, 1:10, and 1:All – were set to represent three scenarios of machine learning from imbalanced positive-negative training data. First, training data with the equal positive and negative data (1:1) were generated per block for each dataset. If the number of negative training pairs in a block is larger than that of positive ones, negative pairs of the amount matched to positive pairs by 1:1 were randomly selected once. If the number of negative pairs in a block is smaller than that of positive pairs, all negative pairs were fed into classifiers. These selection schemes also applied to the 1:10 scenario. Blocks with no negative training pairs were excluded from analysis in all scenarios.

Table 3 summarizes the numbers of blocks that belong to different $R$ ranges in each dataset. In KANG data, for example, 30 blocks (34.88% of all blocks) have ratios of positive-negative training data capped at 1:1. Therefore, when disambiguated for the equal positive and negative (1:1) ratio scenario, all negative pairs in each of 30 blocks will be used for training, while in 56 blocks with $R > 1$, negative pairs will be uniformly sampled to match the size of positive pairs by 1:1 for per-block training.

*Table 3: Summary of Block Distribution per R in GILES, KANG, and TANG Data (R represents the maximum ratio of negative to positive training pairs and the percentage of R blocks over all blocks is reported in parentheses)*

| Data | No. of Instances (train + test) | No. of Authors (train + test) | No. of Blocks (Name Instances ≥ 100) | | | |
|---|---|---|---|---|---|---|
| | | | All | $0 < R \leq 1$ | $1 < R \leq 10$ | $10 < R$ |
| GILES | 5,017 | 480 | 14 | - | 10 (71.43%) | 4 (28.57%) |
| KANG | 13,041 | 2,061 | 86 | 30 (34.88%) | 32 (37.21%) | 24 (27.91%) |
| TANG | 3,984 | 792 | 19 | 6 (31.58%) | 5 (26.32%) | 8 (42.11%) |

The disambiguation accuracy of three classifiers per scenario is presented in Figure 3. Accuracy scores – $b$P, $b$R, and $b$F1 – were averaged over per-block values. According to Figure 3 (a), (d), and (g), increasing the ratios of negative training data from 1:1 to 1:10 to 1:All increased the precision ($b$P) by Random Forest (RF) across three datasets. However, the performances of other two classifiers were not consistent. Logistic Regression (LR) produced slightly higher precision with larger training data in TANG (Fig.3 (g)) but performed slightly worse in GILES (Fig.3 (a)) and KANG (Fig3. (d)). Naïve Bayes (NB) showed a similar pattern: its precision was improved in GILES but decreased in TANG or stalled in KANG. This overall pattern was also observed for recall (see Figure 3 (b), (e), and (h)).

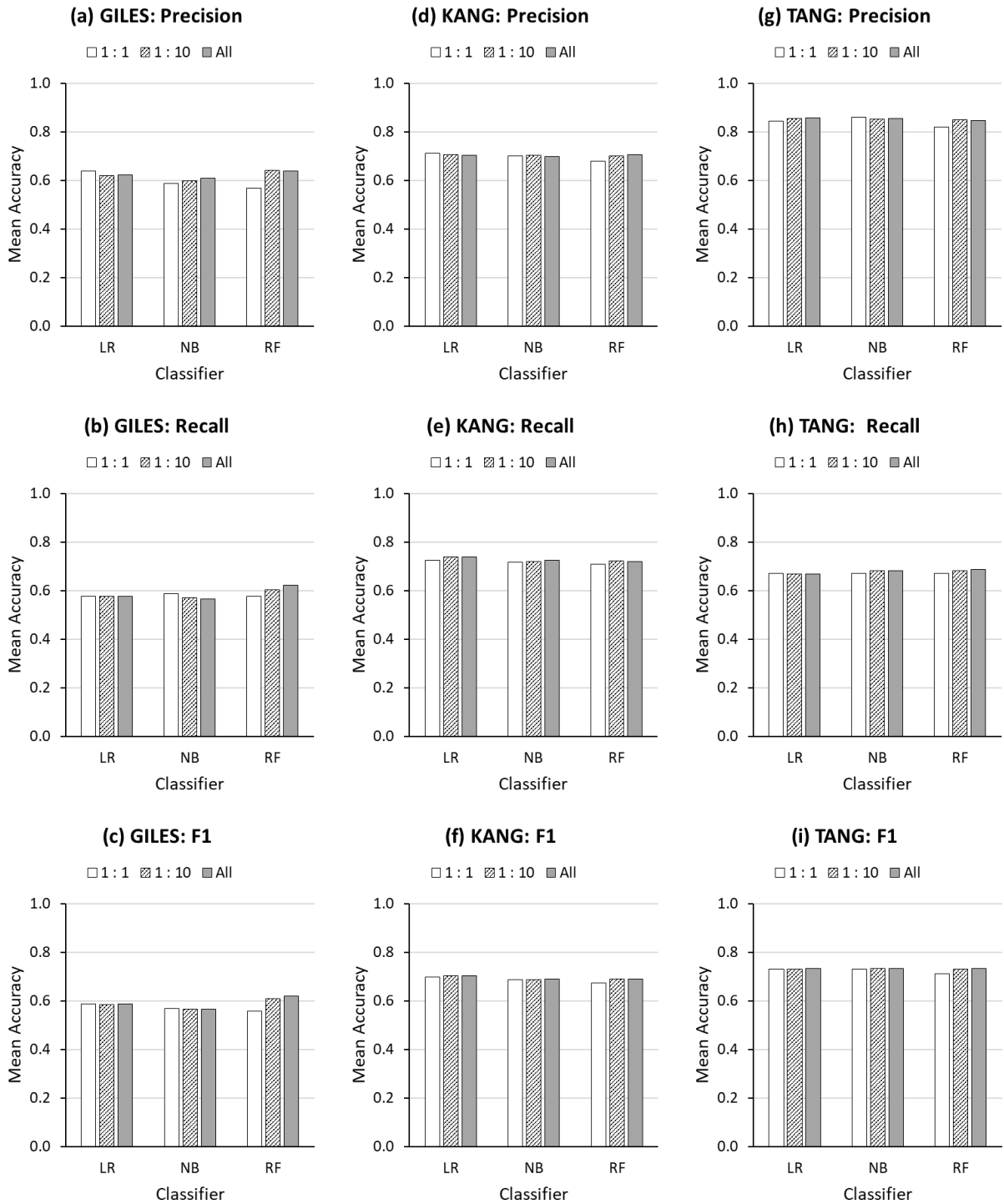

*Figure 3: Mean Accuracy of Author Name Disambiguation per Positive-Negative Training Pair Ratio for GILES, KANG, and TANG Data (y-axes denote mean accuracy scores of B-cubed precision, recall, and F1 measured on test data for 1:1, 1:10, and 1:All positive-negative training ratio scenarios per classifier)*

In contrast to slight variations in precision and recall, their harmonic mean ($b$F1) scores by Logistic Regression and Naïve Bayes were not much differentiated with the increased negative training data. In Figure 3 (c), (f), and (i), specifically, their bar heights are almost the same regardless of positive-negative training data ratios. Random Forest produced a little higher $b$F1 using increased negative training data across three datasets but with a few percent of performance gains. Especially, using the 1:10 ratio of positive-to-negative training data led to the accuracy scores as high as those obtained from all-out use of training data.

These cross-data observations agree with the observations on 10 individual blocks in GILES (Figure 1 and Figure 2). Increased precision with larger negative training data tend to be offset by decreased recall. In addition, due to such a cancelling-out effect between precision and recall, classifiers produced similar $b$F1 scores over different ratios of negative to positive training data. Most importantly, the results aggregated from per-block and cross-data analyses imply that training some classifiers for author name disambiguation may be insensitive to the imbalance of positive and negative training data or sometimes, be adversely affected by increased negative training data.

## Conclusion and Discussion

This paper empirically tested how the ratios of negative to positive training data can affect the performances of machine learning algorithms in disambiguating author names. Using multiple labeled datasets, three classifiers – Logistic Regression, Naïve Bayes, and Random Forest – were trained through two representative features (coauthor names, and title words) extracted from the same labeled data but with various positive-negative data ratios. In terms of the B-cubed precision, recall, and F1 scores, increasing negative training data against positive data improved disambiguation performance by Random Forest, but not much by Logistic Regression and Naïve Bayes classifiers. Even the performance improvement by Random Forest, however, tended to quickly saturate: adding more negative training data beyond certain positive-negative ratios did not contribute much to disambiguation performances. Such findings were tested by repeating 10 times the process of random sampling of negative training sets in this paper. One-standard-deviations of disambiguation outcomes from the repeated sampling were less than 2% of mean values across the sample sizes, which indicates that the trends reported in Figure 1 and 2 are quite robust to the sampling of negative training sets. Such a negligible impact of negative training data on name disambiguation was also confirmed in tests on other two labeled datasets (KANG and TANG in conjunction with GILES in Figure 1 and 2) which are different in size and composition of name ambiguity.

The findings of this study corroborate those of a few studies for predicting pharmacological compounds for virtual screening (Heikamp & Bajorath, 2013; Kurczab et al., 2014). According to the studies, increased negative training data led to the improvement of precision and Mathews Correlation Coefficient (MCC, a measure for balancing precision and recall) and degradation of recall by several algorithms including Random Forest. Beyond the positive-negative data ratio of 1:9 or 1:10, the improvement by added negative training data became negligible, which was also observed in this study through training and testing classifiers on multiple labeled datasets. Another interesting finding is that in this study, adding more negative training data led to deteriorating performance depending on classifiers and accuracy measures. This is in line with the aforementioned virtual screening study reporting degraded recall by added negative training data and also supports partially the PU (positive-unlabeled data) learning approach arguing that "negative training data can be harmful" to machine learning for text classification (Li et al., 2010; Liu et al., 2003).

The results of this study suggest that machine learning algorithms for author name disambiguation can be trained using part of negative training data without much degraded performance. This can be good news for scholars who conduct research on supervised author name disambiguation in which negative training sets generated from large name blocks can skew heavily the distribution of the whole training data. In other words, scholars can use a subset of negative training data for machine learning in author name disambiguation tasks, which can improve the computational efficiency (less amount of negative training data), while similar levels of algorithmic effectiveness are obtainable (similar or slightly degraded disambiguation performance).

Before applying this study's findings to disambiguation tasks, however, several issues must be addressed. First, as reported in Levin et al. (2012), utilizing all available negative training data can be effective in certain circumstances. For example, negative training data size may become impactful to machine learning for author name disambiguation when some features such as author affiliation and publication venue are added or other classifiers than Logistic Regression, Naïve Bayes, and Random Forest are used for model learning.

Second, in-depth research is needed to understand why negative training data do not affect much author name disambiguation. A plausible explanation is that name instances that refer to different authors tend not to share common coauthors and research topics (in terms of title words), producing similarity profiles that are not much discriminative. This means that as information in negative training sets is redundant (i.e., most name pairs in comparison do not share coauthor names and title keywords), the random subsets of negative training sets would contain information similar to that of population. This might explain why the amount of negative training sets does not matter on disambiguation performance of algorithms: whether the amount is small or large, the information an algorithm can be trained is almost the same. Note that the negligible impact of negative training sets was found after the ratio of positive to negative training sets was around 1:10 for the highly imbalanced cases in Figure 1 by Random Forest. This exception might be because Random Forest classifier utilizes the majority voting of outcomes based on sampling of training data (i.e. samples of sampled negative training data) and thus can be more sensitive to sample sizes of negative training data than Logistic Regression and Naïve Bayes. But this conjecture should be investigated in conjunction with why precision gains from increased negative training data tend to be offset by recall losses. Findings from this future investigation can be utilized to increase precision while controlling the adverse impact of negative training data on recall or vice versa. In addition, a more elaborated theory than the aforementioned conjecture would be helpful, which can be modeled by testing the findings of this study on a variety of labeled data under different conditions.

Third, research on the impact of positive training data as well as negative data would be useful. In-depth studies about the relationship between positive and negative training data may help us develop effective training data sampling methods suited for author name disambiguation at scale. Ultimately, this study is expected to motivate scholars to pay more attention to research on supervised machine learning for author name disambiguation with imbalanced (training + test) data in general furthering the scope of studying imbalanced training data.

## Acknowledgments

This work was supported by grants from the National Science Foundation (#1561687 and #1535370), the Alfred P. Sloan Foundation and the Ewing Marion Kauffman Foundation. We would like to thank anonymous reviewers for comments that helped us improve this paper.

# References

Bagga, A., & Baldwin, B. (1998). *Algorithms for scoring coreference chains.* Paper presented at the The first international conference on language resources and evaluation workshop on linguistics coreference.

Barabási, A.-L., Jeong, H., Neda, Z., Ravasz, E., Schubert, A., & Vicsek, T. (2002). Evolution of the social network of scientific collaborations. *Physica a-Statistical Mechanics and Its Applications, 311*(3-4), 590-614.

Bickel, S., Bruckner, M., & Scheffer, T. (2009). Discriminative Learning Under Covariate Shift. *Journal of Machine Learning Research, 10*, 2137-2155.

Chawla, N. V., Bowyer, K. W., Hall, L. O., & Kegelmeyer, W. P. (2002). SMOTE: synthetic minority over-sampling technique. *Journal of artificial intelligence research, 16*, 321-357.

Cota, R. G., Ferreira, A. A., Nascimento, C., Goncalves, M. A., & Laender, A. H. F. (2010). An Unsupervised Heuristic-Based Hierarchical Method for Name Disambiguation in Bibliographic Citations. *Journal of the American Society for Information Science and Technology, 61*(9), 1853-1870.

Fegley, B. D., & Torvik, V. I. (2013). Has large-scale named-entity network analysis been resting on a flawed assumption? *PLOS ONE, 8*(7).

Ferreira, A. A., Goncalves, M. A., & Laender, A. H. F. (2012). A Brief Survey of Automatic Methods for Author Name Disambiguation. *Sigmod Record, 41*(2), 15-26.

Ferreira, A. A., Veloso, A., Goncalves, M. A., & Laender, A. H. F. (2014). Self-Training Author Name Disambiguation for Information Scarce Scenarios. *Journal of the Association for Information Science and Technology, 65*(6), 1257-1278.

Garfield, E. (1969). British quest for uniqueness versus American egocentrism. *Nature, 223*(5207), 763.

Han, H., Giles, L., Zha, H., Li, C., & Tsioutsiouliklis, K. (2004). Two Supervised Learning Approaches for Name Disambiguation in Author Citations. *Jcdl 2004: Proceedings of the Fourth Acm/Ieee Joint Conference on Digital Libraries*, 296-305.

Han, H., Xu, W., Zha, H., & Giles, C. L. (2005). *A hierarchical naive Bayes mixture model for name disambiguation in author citations.* Paper presented at the Proceedings of the 2005 ACM symposium on Applied computing - SAC '05, Santa Fe, New Mexico.

Han, H., Zha, H. Y., & Giles, C. L. (2005). Name disambiguation spectral in author citations using a K-way clustering method. *Proceedings of the 5th Acm/Ieee Joint Conference on Digital Libraries, Proceedings*, 334-343.

He, H., & Garcia, E. A. (2009). Learning from imbalanced data. *IEEE Transactions on Knowledge and Data Engineering, 21*(9), 1263-1284.

Heikamp, K., & Bajorath, J. (2013). Comparison of Confirmed Inactive and Randomly Selected Compounds as Negative Training Examples in Support Vector Machine-Based Virtual Screening. *Journal of Chemical Information and Modeling, 53*(7), 1595-1601.

Kang, I. S., Kim, P., Lee, S., Jung, H., & You, B. J. (2011). Construction of a Large-Scale Test Set for Author Disambiguation. *Information Processing & Management, 47*(3), 452-465.

Kim, J. (2018). Evaluating author name disambiguation for digital libraries: a case of DBLP. *Scientometrics*. doi:10.1007/s11192-018-2824-5

Kim, J., & Diesner, J. (2016). Distortive effects of initial-based name disambiguation on measurements of large-scale coauthorship networks. *Journal of the Association for Information Science and Technology, 67*(6), 1446-1461.

Kim, K., Sefid, A., & Giles, C. L. (2017). Scaling Author Name Disambiguation with CNF Blocking. *arXiv preprint arXiv:1709.09657*.

Kurczab, R., Smusz, S., & Bojarski, A. J. (2014). The influence of negative training set size on machine learning-based virtual screening. *Journal of Cheminformatics, 6*.

Levin, M., Krawczyk, S., Bethard, S., & Jurafsky, D. (2012). Citation-Based Bootstrapping for Large-Scale Author Disambiguation. *Journal of the American Society for Information Science and Technology, 63*(5), 1030-1047.

Li, X.-L., Liu, B., & Ng, S.-K. (2010). *Negative training data can be harmful to text classification.* Paper presented at the Proceedings of the 2010 Conference on Empirical Methods in Natural Language Processing, Cambridge, Massachusetts.

Liu, B., Dai, Y., Li, X., Lee, W. S., & Yu, P. S. (2003). *Building text classifiers using positive and unlabeled examples.* Paper presented at the Data Mining, 2003. ICDM 2003. Third IEEE International Conference on.

Louppe, G., Al-Natsheh, H. T., Susik, M., & Maguire, E. J. (2016). Ethnicity Sensitive Author Disambiguation Using Semi-supervised Learning. *Knowledge Engineering and Semantic Web, Kesw 2016, 649*, 272-287.

Menestrina, D., Whang, S. E., & Garcia-Molina, H. (2010). Evaluating entity resolution results. *Proceedings of the VLDB Endowment, 3*(1-2), 208-219.

Muller, M. C., Reitz, F., & Roy, N. (2017). Data sets for author name disambiguation: An empirical analysis and a new resource. *Scientometrics, 111*(3), 1467-1500.

Newman, M. E. J. (2001). The structure of scientific collaboration networks. *Proceedings of the National Academy of Sciences of the United States of America, 98*(2), 404-409.

Porter, M. (1980). An algorithm for suffix stripping. *Program, 14*(3), 130-137.

Santana, A. F., Goncalves, M. A., Laender, A. H. F., & Ferreira, A. A. (2015). On the Combination of Domain-Specific Heuristics for Author Name Disambiguation: The Nearest Cluster Method. *International Journal on Digital Libraries, 16*(3-4), 229-246.

Schulz, J. (2016). Using Monte Carlo simulations to assess the impact of author name disambiguation quality on different bibliometric analyses. *Scientometrics, 107*(3), 1283-1298.

Shimodaira, H. (2000). Improving predictive inference under covariate shift by weighting the log-likelihood function. *Journal of Statistical Planning and Inference, 90*(2), 227-244.

Shin, D., Kim, T., Choi, J., & Kim, J. (2014). Author Name Disambiguation Using a Graph Model with Node Splitting and Merging Based on Bibliographic Information. *Scientometrics, 100*(1), 15-50.

Smalheiser, N. R., & Torvik, V. I. (2009). Author Name Disambiguation. *Annual Review of Information Science and Technology, 43*, 287-313.

Tang, J., Fong, A. C. M., Wang, B., & Zhang, J. (2012). A Unified Probabilistic Framework for Name Disambiguation in Digital Library. *IEEE Transactions on Knowledge and Data Engineering, 24*(6), 975-987.

Tang, L., & Walsh, J. P. (2010). Bibliometric fingerprints: name disambiguation based on approximate structure equivalence of cognitive maps. Scientometrics, 84(3), 763-784.

Torvik, V. I., & Smalheiser, N. R. (2009). Author name disambiguation in MEDLINE. *Acm Transactions on Knowledge Discovery from Data, 3*(3).

Treeratpituk, P., & Giles, C. L. (2009). Disambiguating Authors in Academic Publications using Random Forests. *JCDL 2009: Proceedings of the 2009 Acm/Ieee Joint Conference on Digital Libraries*, 39-48.

Wang, J., Berzins, K., Hicks, D., Melkers, J., Xiao, F., & Pinheiro, D. (2012). A boosted-trees method for name disambiguation. Scientometrics, 93(2), 391-411.

Wang, X., Tang, J., Cheng, H., & Yu, P. S. (2011). *ADANA: Active Name Disambiguation*. Paper presented at the 2011 IEEE 11th International Conference on Data Mining.

Woods, K. S., Doss, C. C., Bowyer, K. W., Solka, J. L., Priebe, C. E., & KEGELMEYER JR, W. P. (1993). Comparative evaluation of pattern recognition techniques for detection of microcalcifications in mammography. *International Journal of Pattern Recognition and Artificial Intelligence, 7*(06), 1417-1436.